# Design of $\beta$-Ga$_2$O$_3$ Modulation Doped Field Effect Transistors

M.A. Mastro [1], M.J. Tadjer [1], J. Kim [2], F. Ren [3], and S.J. Pearton [4]

[1] U.S. Naval Research Laboratory, Washington, DC 20375, USA
[2] Department of Chemical and Biological Engineering, Korea University, Seoul 02841, Korea
[3] Department of Chemical Engineering, University of Florida, Gainesville, FL 32611, USA
[4] Department of Materials Science and Engineering, University of Florida, Gainesville, FL 32611, USA

**ABSTRACT**

The design of $\beta$-Ga$_2$O$_3$-based modulation doped field effect transistors (MODFETs) is discussed with a focus on the role of self-heating and resultant modification of the electron mobility profile. Temperature- and doping-dependent model of the electron mobility as well as temperature- and orientation-dependent approximations of the thermal conductivity of $\beta$-Ga$_2$O$_3$ are presented. A decrease in drain current was attributed to a position-dependent mobility reduction caused by a coupled self-heating mechanism and a high electric-field mobility reduction mechanism. A simple thermal management solution is presented where heat is extracted through the source contact metal. Additionally, it is shown that an undesired secondary channel can form at the modulation doped layer that is distinguished by an inflection in the transconductance curve.

## I. INTRODUCTION

Monoclinic $\beta$-Ga$_2$O$_3$ based transistors possess fundamental electronic properties that are advantageous for high power devices.[1] A number of these properties derive directly from the wide band-gap of $\beta$-Ga$_2$O$_3$ (E$_g$ = 4.8 eV) including an exceptionally high electric breakdown field (approximately 8 MV/cm).[2] The high breakdown field allows for the design of a lateral transistor with a short channel that can withstand a high critical field. The short channel design lowers the switching loss and, also, lowers the on-resistance, which reduces the conduction loss.[3] Additionally, the short channel allows the high-voltage switch to operate at high frequency (> 1 MHz) and, thus, enables a system design with smaller (size and weight) passive circuit components.[4]

A lateral metal-semiconductor or metal-oxide-semiconductor field effect transistor is typically based on an n-type doped channel. At carrier concentrations greater than 1x10$^{18}$ cm$^{-3}$, the mobility of $\beta$-Ga$_2$O$_3$ is limited to approximately 50 cm$^2$/V·s due to impurity scattering.[5] In contrast, at a carrier concentration of 2.5x10$^{16}$ cm$^{-3}$, a mobility of $\beta$-Ga$_2$O$_3$ of 184 cm$^2$/V·s has been reported – primarily limited by polar optical phonon scattering.[6] This provides the motivation for a modulation doped design where donors within the (Al$_x$Ga$_{1-x}$)$_2$O$_3$ barrier accumulate in a triangular potential well in the undoped Ga$_2$O$_3$ crystal at the (Al$_x$Ga$_{1-x}$)$_2$O$_3$ / Ga$_2$O$_3$ interface. An (Al$_x$Ga$_{1-x}$)$_2$O$_3$ spacer layer physically separates the dopant atoms from the electrons in the conductive channel, which reduces impurity scattering and, thus, improves electron mobility.

With proper design, conduction predominantly occurs via the dense two-dimensional electron gas (2DEG) that forms in the potential well. Nevertheless, a deleterious secondary channel may form at the modulation-doping location in the (Al$_x$Ga$_{1x}$)$_2$O$_3$ barrier. The supplemental to this article provides general design rules for optimal formation of the 2DEG. The supplemental focuses on mitigating three limitations related to current deposition technology involving the aluminum content in (Al$_x$Ga$_{1-x}$)$_2$O$_3$, modulation-doping concentration, and the persistent tail of dopant atoms. Below, it is shown that residual charge in the (Al$_x$Ga$_{1-x}$)$_2$O$_3$ barrier manifests as a secondary inflection in the transconductance curve.

This primary aim of this article is to investigate the coupled relation of on-state self-heating and spatially-dependent mobility reduction, and the resultant alteration of the forward bias response. The mobility decreases significantly at elevated temperature and the low-thermal conductivity of β-Ga$_2$O$_3$ results in self-heating of the device at moderate current density. Approximations to the mobility and thermal conductivity of β-Ga$_2$O$_3$ are presented that can be readily implemented in a TCAD simulation environment. It is shown that self-heating and the electric-field dependence will significantly alter the cross-sectional mobility profile. Significant self-heating can result in a negative differential conductance in the drain current response to increasing drain voltage.

## II. MODELLING

A series of electro-thermal simulations, via the Silvaco TCAD environment, were conducted for a modulation doped (Al$_x$Ga$_{1-x}$)$_2$O$_3$ / Ga$_2$O$_3$ structure to provide insight into the device current-voltage relationships as well as the cross-sectional profiles of current density, mobility, and temperature.

### A. Thermal Conductivity and Mobility Models

Accurate description of the temperature dependent parameters is necessary for simulation of a transistor based on Ga$_2$O$_3$, which suffers from self-heating related to its low thermal conductivity. For application to a TCAD simulation, the thermal conductivity model given in Mu et al.[7] was fit to a power-law thermal model given by

$$\kappa(T) = \kappa_{300}\left(\frac{T}{300}\right)^\alpha, \quad (1)$$

with the parameters listed in Table 1.

|  |  | $\kappa_{300}$ [$W/cm \cdot K$] | $\alpha$ |
|---|---|---|---|
| Ga$_2$O$_3$ | [100] | 0.111 | -1.33487 |
|  | [010] | 0.155 | -1.25526 |
|  | [001] | 0.153 | -1.31247 |
| AlGaO$_3$ | [100] | 0.198 | -1.62331 |
|  | [010] | 0.25 | -1.48543 |
|  | [001] | 0.304 | -1.61316 |
| Heat-Sink |  | 1.8 | -1.3 |

TABLE 1. Orientation dependent model parameters for thermal conductivity for Ga$_2$O$_3$ and AlGaO$_3$, and a heat-sink material comparable to ceramic poly-AlN.

Ma et al. developed a model of mobility of Ga$_2$O$_3$, that accounted for several scattering mechanisms.[8] The Ma mobility model, based on the Boltzmann transport equation via the relaxation-time approximation solution, requires two separate integrations. To introduce a temperature and donor dependent model into the device simulation, the Boltzmann model of Ma et al. was fit to the Arora low-field mobility model[9] with a structure in the Silvaco TCAD environment given by

$$\mu_n = \mu_1\left(\frac{T}{300}\right)^\alpha + \frac{\mu_2\left(\frac{T}{300}\right)^\beta}{1+\frac{N}{N_{crit}\left(\frac{T}{300}\right)^\gamma}}, \quad (2)$$

with fitting parameters given by $\mu_1$ = 1.67938 cm$^2$/V·s, $\mu_2$ = 264.557 cm$^2$/V·s, $\alpha$= -0.116259, $\beta$ = -2.29257, $\gamma$ = 2.64426, and $N_{crit}$ = 4.29447×10$^{18}$ cm$^{-3}$.

Figure 1(a) displays the calculated mobility of the Arora model compared to the calculated Ma model and the specific scattering mechanisms defined in the Ma model. Figure 1(b) and 1(c) displays the Arora calculated mobility as a function of dopant density and temperature. The Arora model replicates the Ma model, which predicts a transition from polar optical phonon scattering as the limiting mechanism at a low donor level to impurity scattering as the dominant mechanism limiting the mobility, above 200 °K, at a high donor level.

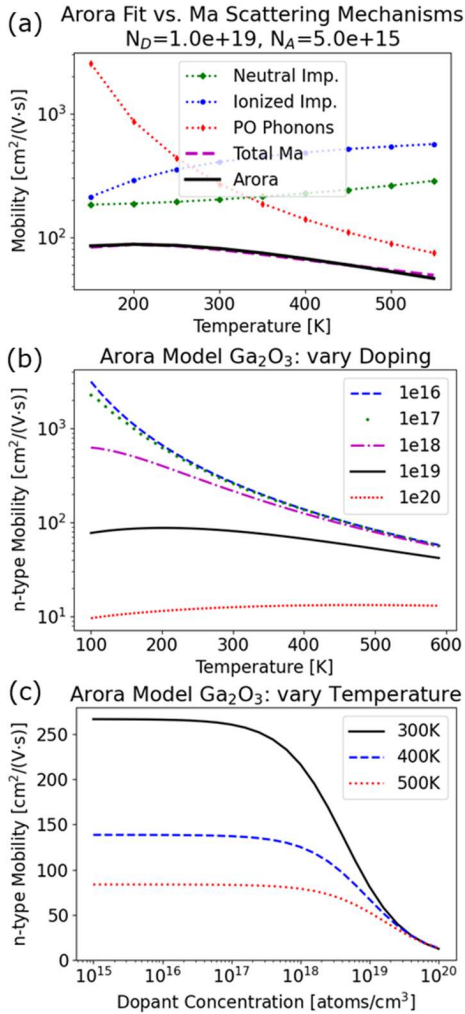

FIG. 1. (Color online) (a) Comparison of fitted Arora model to Ma mobility model with its relevant scattering mechanisms. Mobility of β-Ga$_2$O$_3$ in the Arora model as a function of (b) temperature and (c) donor concentration.

## B. Device Structure

The base simulation structure consists of 20 nm ALD Al$_2$O$_3$, 30nm (Al$_{0.25}$Ga$_{0.75}$)$_2$O$_3$ with a background donor concentration of 1×10$^{17}$ cm$^{-3}$ and modulation doped at 4×10$^{19}$ cm$^{-3}$ with a Gaussian distribution with a characteristic length of 0.5 nm that peaks at 4nm from the layer bottom, 200 nm Ga$_2$O$_3$ with a background donor concentration of 1×10$^{15}$ cm$^{-3}$, and a 200 μm ($\bar{2}$01) Ga$_2$O$_3$:Fe substrate with a thermal boundary resistance of 0.001 cm$^2$·K/W. The source and drain contacts are 50 μm in length and composed of 20 nm Ti and 300 nm Au. The gate contact is 5 μm in length and composed of 20 nm Ni and 300 nm Au. The source to gate spacing is 3 μm and the gate to drain spacing is 12 μm. The remaining structure consists of Si$_3$N$_4$.

As shown in Fig. 2 a heatsink, with a temperature dependent thermal conductivity similar to ceramic poly-AlN, is optionally added. The heatsink is 50 μm in length and ends 5 μm from the edge of the source metal. The heatsink is separated by 50 nm of Si$_3$N$_4$ from the source metal. The heatsink has a thermal boundary resistance of 0.001 cm$^2$·K/W and the Ga$_2$O$_3$:Fe substrate is thermally isolated.

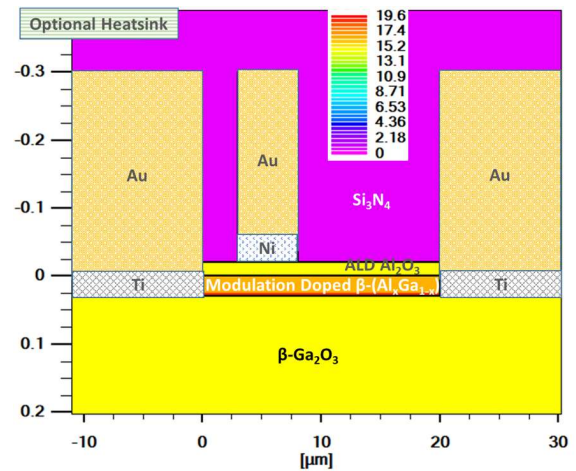

FIG. 2. (Color online) Components and log donor concentration [cm$^{-3}$] profile of base simulation structure. The MODFET is designed to have the primary path of current flow in the triangular potential well formed in the first few nanometers in the undoped Ga$_2$O$_3$ layer below the (Al$_{0.25}$Ga$_{0.75}$)$_2$O$_3$ / Ga$_2$O$_3$ interface.

## III. RESULTS

### A. Electro-Thermal Simulation with Thermal Contact to Substrate Base

A simulation of the device structure was conducted, with a thermal contact to the substrate base, as described in the previous section. The drain current and peak temperature as a function of forward drain bias are displayed in Fig. 3(a). At a drain bias above 3 V, the temperature of the device rapidly increases. Additionally, above 7 V, a negative differential conductivity is observed. The increase in temperature relative to the increase in power corresponds to a device thermal resistance ($R_{th}$ = $\Delta T/P_D$) of 167 mm·K°/W at 10 V (and 359 °K).

Figure 3(b) displays the drain current and peak temperature as a function of gate bias for drain voltage of 5 V and 10 V. At a drain voltage of 5 V and 10 V, the peak device temperature plateaus at approximately 325 °K and 385 °K, respectively. At a large positive gate bias, the gate minimally impacts the channel, which behaves similar to a resistor. Moving from a drain voltage of 5 V to 10 V corresponds to a 47% increase in drain current but a 132% increase in temperature.

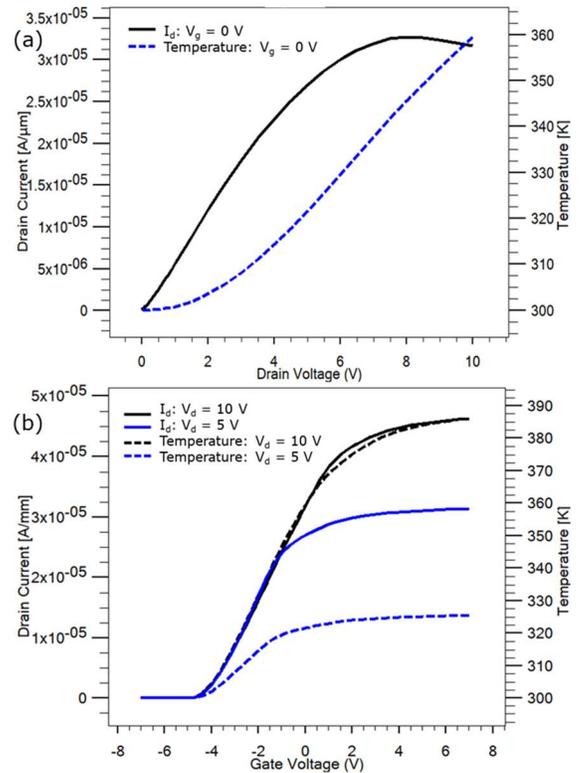

FIG. 3. (Color online) Drain current and peak temperature as a function of (a) drain voltage at zero gate bias and (b) gate bias at a drain voltage of 5 V and 10 V. The temperature vs. drain bias in (a) displays a slope of 8.2 K°/V for $V_d$ greater than 5 V.

Figure 4 compares the log current density, x-component (direction from source to drain) of the mobility, and temperature at a fixed drain current (of $V_d$ = 10 V). The left-column corresponds to the near complete depletion of the channel (at $V_g$ = -4 V), center-column corresponds to the point of initial depletion (starting at the drain side of the gate edge) in the channel (at $V_g$ = 0 V), and right-column is the transitional point of depletion of the (Al$_{0.25}$Ga$_{0.75}$)$_2$O$_3$ barrier (at $V_g$ = 2 V) in Fig. 4.

In Fig. 4(d) and 4(e), the reduction in mobility at the drain side of the gate edge is caused by the peak in x-component of the electric field as predicted by the high-field dependent mobility model. This causes a local increase in resistance, which similarly causes an increase in Joule heating and a corresponding peak in temperature at the drain side of the gate edge. In Fig. 4(a) (at $V_g$ = -4 V) it is seen that the total current is low, thus, the Joule heating throughout the channel is low and the overall temperature increase in the structure is minimal as observable in Fig. 4(g).

In contrast, in Fig. 4(b) (at $V_g$ = 0 V) a significant current is flowing in the channel, which causes Joule heating along the length of the channel as well as a relatively higher increase in heating at the drain edge of the gate (due to the high electric-field induced mobility reduction). This generated heat sharply increases the temperature at the drain side of the gate edge in the channel (Fig. 4(h)). A temperature gradient exists moving away from the gate edge but the temperature in the surrounding structure is sufficient to cause an appreciable reduction in the mobility in the corresponding region (Fig. 4(e)).

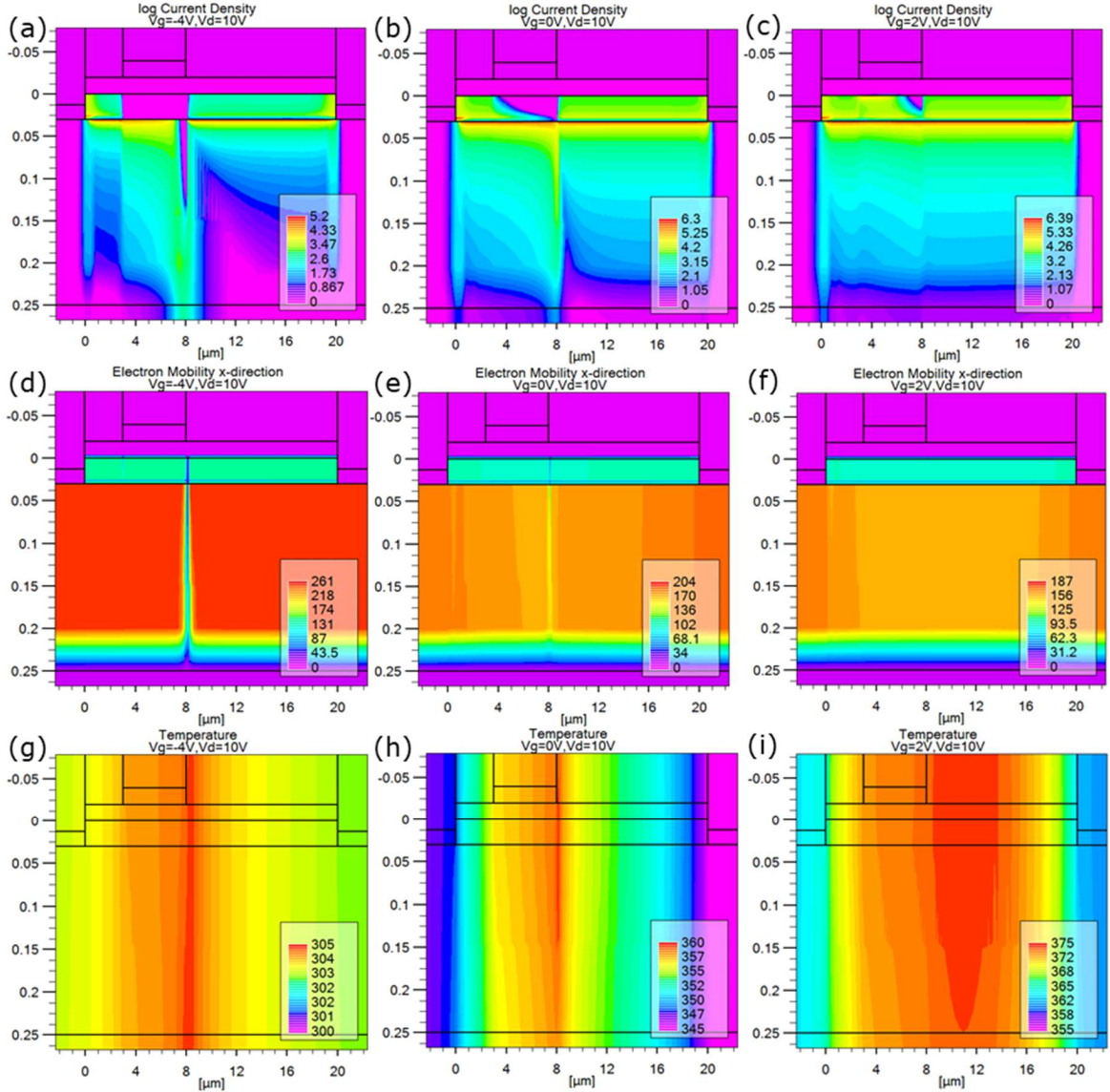

FIG. 4. (Color online) Cross-sections with $V_d$ at 10 V and with $V_g$ at -4 V (a, d, and g), with $V_g$ at 0 V (b, e, and h), and with $V_g$ at 2 V (c, f, and i) for (a, b, c) log current density [A/cm$^2$], (d, e, f) x-component of electron mobility [cm$^2$/V·s], and (g, h, i) lattice temperature [°K]. The left-column corresponds to the near complete depletion of the channel (at $V_g$ = -4 V), which greatly reduces (a) current flow; a high electric field reduces the (d) mobility at the gate edge, however, the overall current flow is low, which limits Joule heating. The center-column corresponds to the point of initial depletion (starting at the drain side of the gate edge) of the channel (at $V_g$ = 0 V) with (b) significant current flowing along the channel; (e) the sharp reduction in mobility at the drain side of the gate edge is caused by the peak in electric field and the coupled self-heating effect on (h) temperature. The right-column corresponds to the transitional point of depletion of the (Al$_{0.25}$Ga$_{0.75}$)$_2$O$_3$ barrier (at $V_g$ = 2 V) with (c) large current flowing through the channel with little interaction with the electric field from the gate; (i) a broad heating and temperature profile occurs owing to a relatively uniform Joule heating along the length of the channel.

The transitional point of depletion of the $(Al_{0.25}Ga_{0.75})_2O_3$ barrier (at $V_g$ = 2 V) reveals a small yet discernable secondary current flow in $(Al_{0.25}Ga_{0.75})_2O_3$ barrier near the location of the modulation doping (Fig. 4(c)). Additionally, at this gate bias, the electric field from the gate has essential no interaction with the primary channel. Hence, the potential well below the $(Al_{0.25}Ga_{0.75})_2O_3$ / $Ga_2O_3$ interface effectively acts as a high-conductivity resistor. The manifests as heating along the entire channel that creates a non-localized heating profile (Fig. 4(i)) and a relatively uniform reduction in mobility in the $Ga_2O_3$ (Fig. 4(f)).

## B. *Comparison to Isothermal Simulation*

A simulation was conducted of the same device structure except thermal heating is neglected, that is, the device is held at 300 °K. Examination of the mobility cross-section, in Fig. 5(b), reveals a sharp decrease in mobility at the drain side of the gate edge – and a smaller decrease at the edge of the source contacts. This decrease in mobility is due to local peak in electric field. Overall, the iso-thermal device has a high mobility along the channel, which corresponds to high current density in Fig 5(a) as compared the electro-thermal simulation in Fig 4(b).

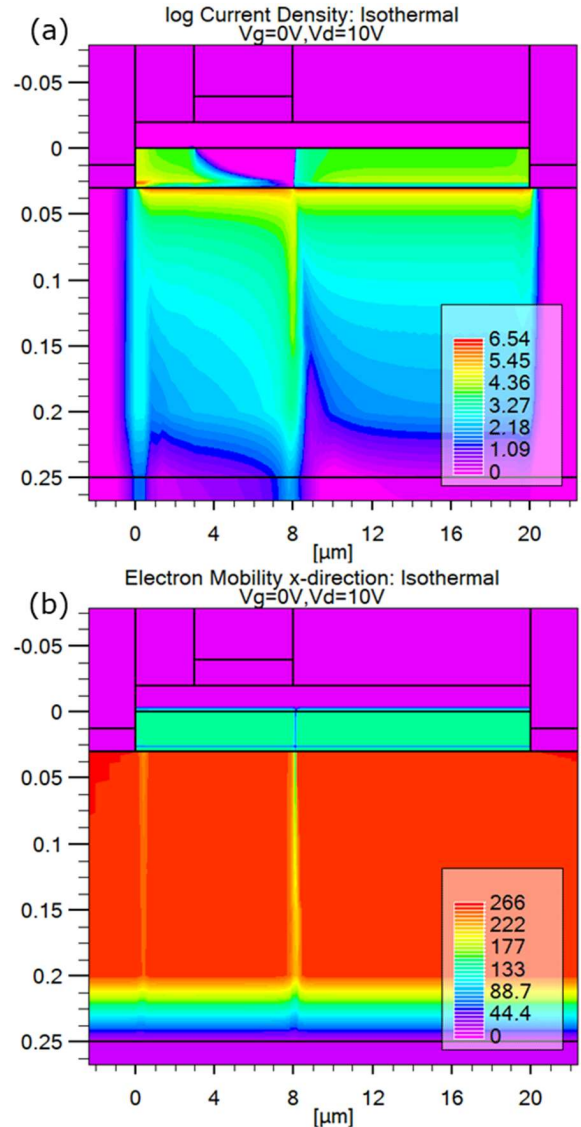

FIG. 5. (Color online) Cross-sections from an iso-thermal simulation at 300 °K for $V_d$ at 10 V for (a) log current density [A/cm$^2$] and (b) x-component of electron mobility [cm$^2$/V·s]. As is common in transistors without field plates, there is a high electric field at the drain side of the gate edge, which results in (b) a local reduction in the x-component of electron mobility.

Figure 6 provides a comparison of the electro-thermal simulations to the equivalent isothermal simulations where the lattice temperature is fixed at 300 °K. The drain current at a drain voltage less than 3 V is similar in the thermal and iso-thermal simulations. This indicates that current density is sufficiently low such that the Joule heat generated can be effectively conducted to and dissipated at the base of the substrate. Additionally, although self-heating is minimal, it is clear the drain voltage is shifting the onset of saturation in drain current – for this particular device geometry and structure.

The drain current and transconductance at a drain voltage of 5 V show a small deviation in the thermal and iso-thermal simulations for a gate bias greater than -1 V. At this transition point, the current density is sufficiently high that most but not all the Joule heat generated can be effectively dissipated at the base of the substrate. For a drain voltage of 5 V, an inflection is present in the transconductance curve (Fig. 6(b)) at a gate bias of 1 V for both the thermal and iso-thermal simulation. This inflection corresponds to the point where the gate voltage is sufficiently positive such that charge, located at modulation doping position in the $(Al_{0.25}Ga_{0.75})_2O_3$ barrier, is no longer depleted. For a drain voltage of 10 V, a similar inflection is present in the transconductance curve (Fig. 6(b)) at a gate bias of 2 V.

As discussed in the previous section, at a drain voltage of 10 V, self-heating can raise the temperature of the lattice, which significantly decreases the electron mobility in and near the channel. This increases the resistance of the device, which is evident in the large difference in the thermal and iso-thermal simulations in Fig. 6.

Figure 6(a) shows a negative differential conductivity for a drain voltage above 7 V with a gate bias of -2 V and 0 V. For this particular device structure, negative differential conductivity is not seen for the isothermal simulations. Here, two coupled mechanisms contribute to the negative differential conductivity; specifically, the sharp reduction in mobility at the drain side of the gate edge caused by the peak in electric field, and the overall reduction in mobility along the entire length of the channel caused by the temperature rise related to self-heating.

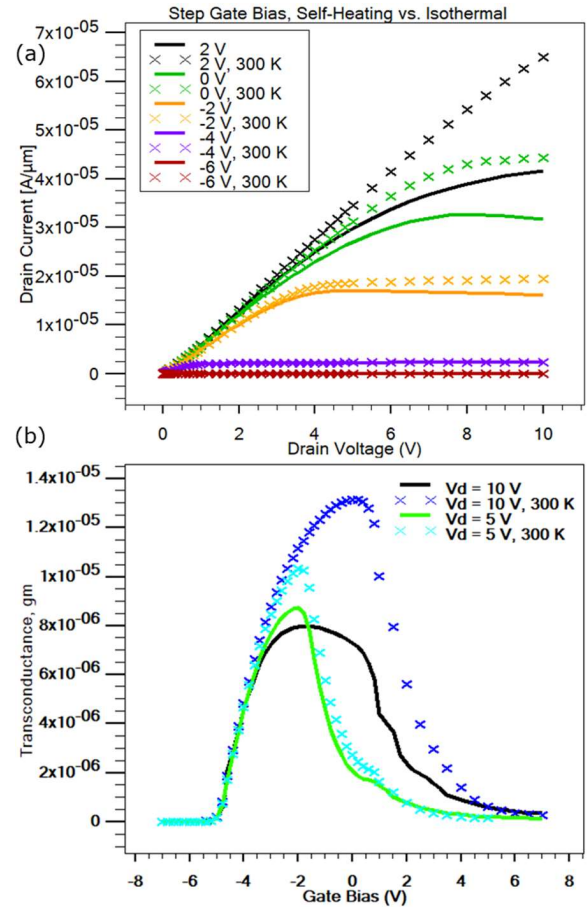

FIG. 6. (Color online) Comparison of electro-thermal simulation with self-heating vs. iso-thermal simulation at 300 °K. (a) Drain current as a function of drain voltage for series of 2 V steps in gate voltage. (b) Transconductance as a function of gate bias at a drain voltage of 5 V and 10 V. Current saturation due to the high-field mobility reduction effect and global heating mechanisms are inter-related. Still, a significant reduction is observed in drain current owing to the reduction in mobility due to self-heating compared to the iso-thermal simulation.

### C. Heat Removal Through the Source Metal

A simple improvement to the thermal design is to place a heatsink material on the surface of the structure near but offset from the source metal. Most transistor test structures have a large source pad above the source region; that is, the source contact is not accessed via a metal finger. As described above, the device surface is covered with a $Si_3N_4$ layer. A 50 nm $Si_3N_4$ layer is between the top of the source metal and the heatsink material, and the heatsink material is offset 5 μm from the edge of the source contact. To further simplify the analysis, no heat exits the bottom of the substrate.

Figure 7(a), shows that the drain current saturates at a drain bias greater than 7V although a slight negative differential conductivity is evident at a drain bias of 10 V. Examination of the temperature cross-section, in Fig. 7(b), reveals the expected peak at the drain side of the gate edge – and an overall gradient towards the heat sink located near the source contacts. The increase in temperature relative to the increase in power corresponds to a thermal resistance, $R_{th}$, of 114 mm·K°/W at 10 V (and 327 °K). The thermal resistance (in Fig. 7) for the source metal thermal contact is 41% the thermal resistance for the bottom substrate thermal contact (in Fig. 3).

A similar thermal resistance of the overall device is found in a simulation (not shown) for a device formed on a 20 μm substrate with a bottom thermal contact. This result is reasonable as the 20 μm distance to the substrate base is similar to the distance from the drain edge of the gate to the source contact. Hence, thinning the substrate is another approach; however, the propensity for β-$Ga_2O_3$ to cleave discourages a post-growth wafer thinning approach. Still, several works have demonstrated devices processed on exfoliated $Ga_2O_3$ films with a thickness typically less than 10 μm.[10-13]

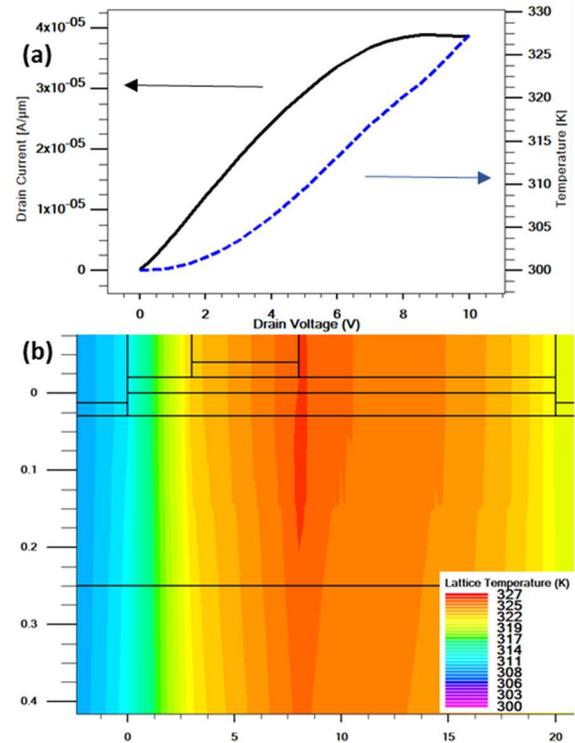

FIG. 7. (Color online) Simulation with heat extraction via source metal. (a) Drain current and temperature as a function of drain voltage at zero gate bias. (b) Temperature [°K] cross-sections for $V_d$ at 10 V. The temperature vs. drain bias in (a) displays a slope of 3.4 K°/V for $V_d$ greater than 5 V.

## IV. DISCUSSION

### A.  Self-Heating and Mobility

Thermal management is an important design consideration in all semiconductor power devices.[4,14] The simulations in this work show that self-heating owing to the low thermal conductivity of β-$Ga_2O_3$ dramatically lowers the mobility in the channel region, which increases the on-state resistance at moderate current densities. The low thermal conductivity of β-$Ga_2O_3$ necessitates advanced thermal management solutions for operation at even moderate power levels.[15,16] Candidate solutions include flip-chip designs where heat is extracted globally from the top surface as well as more nuanced modification of the gate region to enhance local heat extraction at or near the gate.[17-19] Naturally, thermal mapping of the device surface or depth directly measures the thermal response of the device.[20] Nevertheless, these measurements can be obscured by the contact structure. Additionally, a thermo-reflectance imaging measurement is complicated by the requirement for illumination greater than the wide bandgap of β-Ga$_2$O$_3$. As outlined in this paper, understanding the relation of self-heating on the response of the drain current to applied gate bias and drain voltage provides an additional assessment of the thermal management in the device structure.

### B. Modulation Doping Considerations of β-Ga$_2$O$_3$ MODFET

The presence of residual charge at the point of modulation doping is undesired as this creates a secondary channel that hampers the ability to modulate the primary channel. This charge at the modulation doping point can be confirmed by an inflection in the transconductance curve. Specifically, as this charge is spatially closer to the gate metal, this inflection occurs at more positive gate bias relative to the primary channel, which is deeper in the structure.

The supplemental to this article presents design rules for the modulation-doped structure with a goal of maximizing the concentration of electrons in the potential well while minimizing the low-mobility secondary channel at the modulation-doped layer. An emphasis is placed on mitigating three limitations related to the deposition of β-Ga$_2$O$_3$; specifically, the inability to deposit high-quality β-(Al$_x$Ga$_{1-x}$)$_2$O$_3$ with an aluminum mole fraction greater than approximately 0.25, the difficulty to deposit a high concentration of dopant atoms without the formation of its native oxide, and the unintentional dopant atom exponential decay tail. Simulations for an aluminum mole fraction of 0.25 show that proper selection of the barrier thickness and modulation-doping level, including accounting for any asymmetric broadening, can yield a dense electron channel confined primarily to the potential well at the (Al$_x$Ga$_{1-x}$)$_2$O$_3$ / Ga$_2$O$_3$ interface.

## V. SUMMARY AND CONCLUSIONS

The wide bandgap and related high critical breakdown field of β-Ga$_2$O$_3$ enables a high-voltage lateral transistor designed for low on-resistance as well as operation at high-frequency, which reduces the size and weight of the passive circuit elements in the overall system.[21,22] In this work an accurate approximation for mobility as a function of temperature and dopant concentration is presented and employed in an electro-thermal simulation of an (Al$_x$Ga$_{1-x}$)$_2$O$_3$ /Ga$_2$O$_3$ MODFET. A decrease in drain current was attributed to a position-dependent mobility reduction caused by a coupled self-heating mechanism and a high electric-field mobility reduction mechanism. Thermal management is a key design issue in all power semiconductor devices. The low thermal conductivity of Ga$_2$O$_3$ hampers any design where heat must flow over any appreciable thickness of Ga$_2$O$_3$. A heatsink via the source contact is discussed and shown to reduce the temperature of the active device channel.

ACKNOWLEDGMENTS
The work at NRL was partially supported by DTRA Grant No. HDTRA1-17-1-0011 (Jacob Calkins, monitor) and the Office of Naval Research. The work at UF is partially supported by HDTRA1-17-1-0011 and HDTRA1-20-2-0002. The content of the information does not necessarily reflect the position or the policy of the federal government, and no official endorsement should be inferred. The work at Korea University was supported by the Korea Institute of Energy Technology Evaluation and Planning (20172010104830) and the National Research Foundation of Korea (2020M3H4A3081799).

DATA AVAILABILITY

The data that support the findings of this study are available from the corresponding author upon reasonable request.

# Design of $Ga_2O_3$ Modulation Doped Field Effect Transistors:
# Supplemental Information

M.A. Mastro [1], M.J. Tadjer [1], J. Kim [2], F. Ren [3], and S.J. Pearton [4]

[1] U.S. Naval Research Laboratory, Washington, DC 20375, USA
[2] Department of Chemical and Biological Engineering, Korea University, Seoul 02841, Korea
[3] Department of Chemical Engineering, University of Florida, Gainesville, FL 32611, USA
[4] Department of Materials Science and Engineering, University of Florida, Gainesville, FL 32611, USA

## I. BACKGROUND

This supplemental provides general design rules for a $Ga_2O_3$-based MODFET with a focus on mitigating limitations in modulation-doping concentration, aluminum content in the $(Al_xGa_{1-x})_2O_3$ barrier, and the exponential tail of dopant atoms. This analysis is based on a one-dimensional coupled Schrodinger-Poisson calculation.

Practically, the deposition technology for β-$Ga_2O_3$-based MODFET is hindered by three issues; namely, the inability to deposit high-quality β-$(Al_xGa_{1-x})_2O_3$ with an aluminum mole fraction greater than approximately 0.25, the difficulty to deposit a high concentration of dopant atoms without the formation of its native oxide, and the persistent tail or memory-effect of dopant atoms.

Current deposition technology is unable to consistently deposit high-quality alloys of β-$(Al_xGa_{1-x})_2O_3$ with a mole fraction, x, greater than 0.25. An early study by Hill et al. into the equilibrium diagram of $Al_2O_3$ in β-$Ga_2O_3$ found the presence of a stable phase of $AlGaO_3$ (i.e., x = 0.5).[1] This work reported that this phase required a temperature of 800 °C, which may preclude lower temperature growth techniques such as MBE. Examination of the enthalpy of formation model of Peelaers et al. suggests that the monoclinic phase is stable for x < 0.71.[2] The expansion the available alloy range for epitaxy of $(Al_xGa_{1-x})_2O_3$ is an important area of future research particularly for the development of the β-$Ga_2O_3$ based MODFET and its variants.[3,4]

The demonstration of modulated doping of atoms was a major advancement in the deposition technology for compound semiconductors and enabled the development of the high electron mobility transistor (HEMT).[5] It was subsequently found that minimizing the deposition temperature for a given diffusion profile, e.g., Si dopant atom in $Al_xGa_{1-x}As$ deposited by MBE at a temperature less than 550 °C, was necessary to form a dopant layer with a delta-function distribution whose width is narrower than the electron de Broglie wavelength.[3,4]

The delta doping of Si into $Ga_2O_3$ is hampered by the presence of an oxidizing environment either during MOCVD or from the oxygen atoms inherent to the neighboring lattice, which is clearly independent of deposition technology. In general, Si readily reacts to form its native oxide, which will prevent the Si atoms from behaving as a dopant in the $Ga_2O_3$ crystal. The alternative is to form a thicker moderately-doped $Ga_2O_3$ layer rather than a distinct delta-doped sheet. Baldini et al. found in MOCVD efficient activation of the Si dopant to produce free carriers in the range of $1 \times 10^{17}$ to $8 \times 10^{19}$ cm$^{-3}$ in β-$Ga_2O_3$ films. For the alternative dopant choice of the Sn, incorporation was hampered above a concentration of $1 \times 10^{19}$ cm$^{-3}$.[6]

In the recent development of MBE of Ga$_2$O$_3$, a persistent impurity tail is attributed to dopants and impurities riding on the growing surface.[7] Bhattacharyya et al. found a similar Si memory effect for doping of Ga$_2$O$_3$ with a MOCVD process.[8] Specifically, Bhattacharyya et al. reported a Si dopant decay slope of 50 nm/decade at 810 °C, 20 nm/decade at 700 °C and 5 nm/decade at 600 °C in the subsequently deposited Ga$_2$O$_3$ layer.[8] This is undesirable as this extended density of charge in the barrier layer can form a low-mobility secondary channel that is deleterious for transistor operation.

This supplemental provides general design rules for a Ga$_2$O$_3$-based MODFET with a focus on maximizing the density of carriers in the triangular potential well formed at the (Al$_x$Ga$_{1-x}$)$_2$O$_3$ / Ga$_2$O$_3$ interface - while eliminating the formation of a secondary channel at the modulation doped layer.

## II. MODELLING

A series of simulations based on a coupled Schrodinger-Poisson one-dimensional calculation were conducted for a (Al$_x$Ga$_{1-x}$)$_2$O$_3$ / Ga$_2$O$_3$ structure to provide insight into the distribution of carriers, conduction band structure, and the wave function in the potential well.[9, 10] A custom numerical fitting and comparison procedure is employed to determine the best fit lines from a range of polynomial, exponential, and logarithmic functions. A custom numerical optimization procedure based on gradient descent is employed to find locally optimal device structures.

## III. RESULTS

### A. *Comparison of Al mole fraction in the barrier layer*

Figure S1(a) displays a one-dimensional profile of the conduction band, electron density, doping level, and wave function for an (Al$_{0.5}$Ga$_{0.5}$)$_2$O$_3$ / Ga$_2$O$_3$ modulation-doped structure, that is, the barrier Al mole fraction is 0.5. As expected, a triangular potential well forms at the (Al$_{0.5}$Ga$_{0.5}$)$_2$O$_3$ / Ga$_2$O$_3$ interface. The (Al$_{0.5}$Ga$_{0.5}$)$_2$O$_3$ barrier is composed of a surface barrier, modulation-doped layer, and spacer layer. The 30 nm (Al$_{0.5}$Ga$_{0.5}$)$_2$O$_3$, surface barrier serves to separate the active structure from depletion from the (q$\phi_B$ = 1.4 eV) Schottky barrier metal. A 0.5 nm (Al$_{0.5}$Ga$_{0.5}$)$_2$O$_3$ layer is modulation-doped at a concentration of 2x10$^{20}$ cm$^{-3}$ to provide a source of carriers (1x10$^{13}$ cm$^{-2}$ total) that accumulate in the potential well. A 3 nm (Al$_{0.5}$Ga$_{0.5}$)$_2$O$_3$ spacer layer is necessary physical separate the dopant atoms and their concomitant impurity scattering centers from the electrons in the potential well. The background density of active electrons, that is the unintentional doping (UID), in all layers is set at 5x10$^{15}$ cm$^{-3}$.

In contrast, Figure S1(b) displays the same profiles for an (Al$_{0.25}$Ga$_{0.75}$)$_2$O$_3$ / Ga$_2$O$_3$ modulation-doped structure, that is, the Al mole fraction is reduced to 0.25. The clear effect is that a secondary channel forms at the modulation-doped layer. The primary channel concentration in the triangular potential well is only slightly reduced. At the modulation-doped layer, the V-shaped conduction band extends below the Fermi level in the barrier, which results in the formation of a secondary conductive channel.

Figure S1(c) compares the primary channel concentration and the ratio of primary-to-secondary channel concentration for a range of Al mole fraction. Increasing the aluminum composition in the barrier increases the conduction band offset between (Al$_x$Ga$_{1-x}$)$_2$O$_3$ and Ga$_2$O$_3$, which increases the confinement of carriers in the potential well.

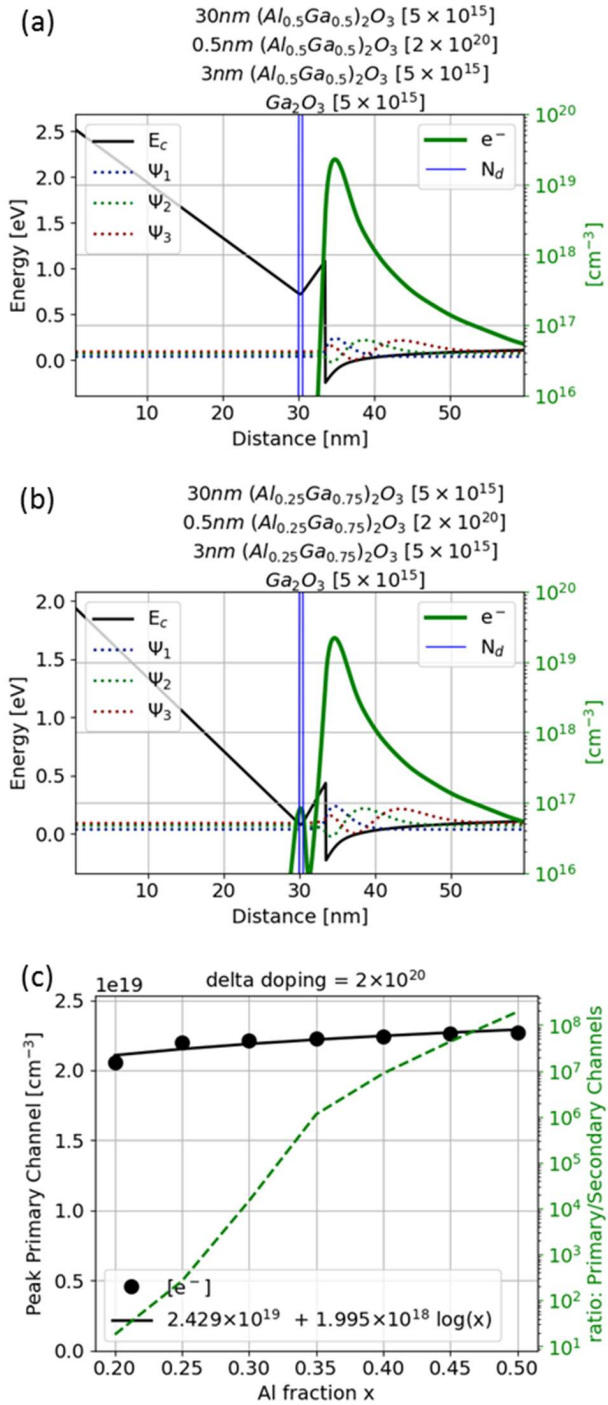

FIG. S1. Variation in Al mole fraction for a modulation-doped β-$(Al_xGa_{1-x})_2O_3$ / β-$Ga_2O_3$ structure where the primary conductive channel forms at the triangular potential well. (a, b) Calculation of energy band, electron concentration, and electron wave function for a structure with a barrier layer composed of $(Al_{0.5}Ga_{0.5})_2O_3$ and $(Al_{0.25}Ga_{0.75})_2O_3$, respectively, on a thick $Ga_2O_3$ layer. (c) Comparison of primary channel concentration and the ratio of primary-to-secondary channel for a range of Al mole fraction. For low Al mole fraction in the barrier layer, a secondary channel forms at the modulation-doping layer.

## B. Modulation-doping

Given the current limitations in maximum aluminum incorporation of approximately 0.25, it is important to examine the $(Al_{0.25}Ga_{0.75})_2O_3$ / $Ga_2O_3$ modulation-doped structure. Figure S2(a) displays a one-dimensional profile for this structure with a modulation-doping concentration of $1 \times 10^{20}$ cm$^{-3}$ for a total sheet concentration of $5 \times 10^{12}$ cm$^{-2}$. A moderate density of electrons accumulates in the triangular potential well at the $(Al_{0.25}Ga_{0.75})_2O_3$ / $Ga_2O_3$ interface.

In contrast, Fig. S2(b) displays the same profiles for an $(Al_{0.25}Ga_{0.75})_2O_3$ / $Ga_2O_3$ modulation-doped structure with a modulation-doping concentration of $4 \times 10^{20}$ cm$^{-3}$ for a total sheet concentration of $2 \times 10^{13}$ cm$^{-2}$. The increase in modulation-doping creates a secondary channel at the modulation-doped layer.

Figure S2(c) compares the primary channel concentration and the ratio of primary to secondary channel concentration for a range of delta doping. For this $(Al_{0.25}Ga_{0.75})_2O_3$ barrier, that is with an Al mole fraction of 0.25, there is a trade-off where increasing the modulation-doping increases the primary channel concentration but also decreases the fraction of carriers in the primary channel relative to the secondary channel at the modulation-doped layer.

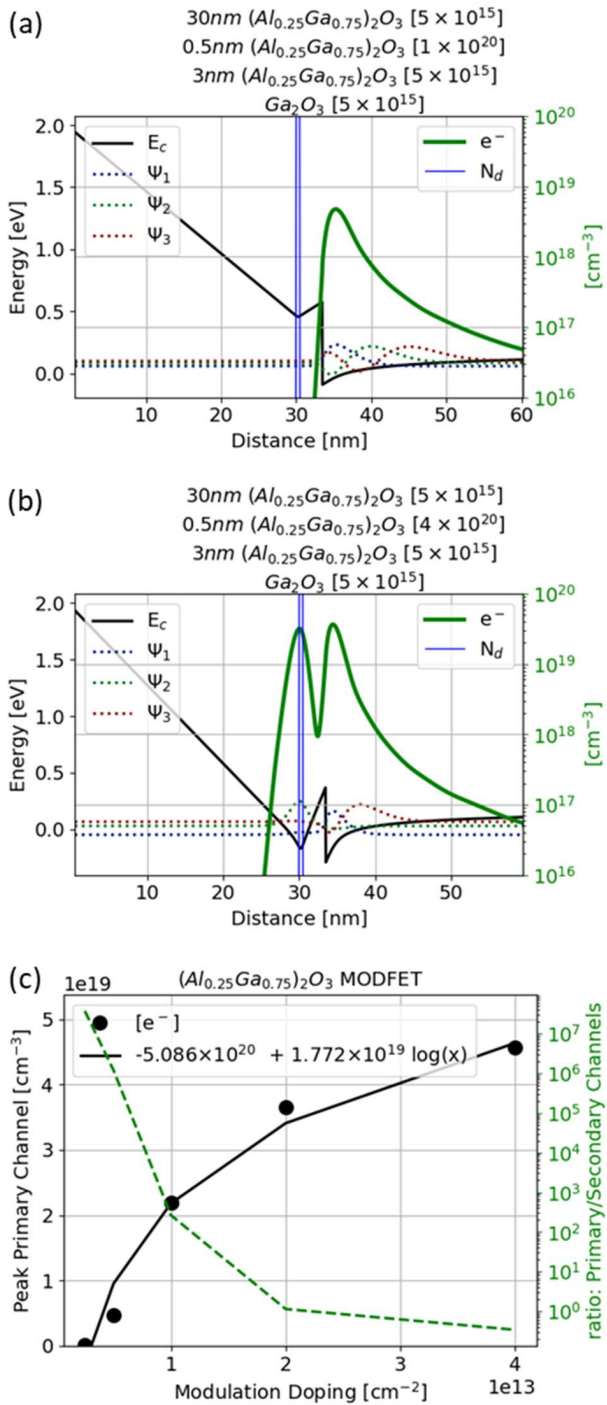

FIG. S2. Variation in modulation-doping concentration for a β-$(Al_{0.25}Ga_{0.75})_2O_3$ / $Ga_2O_3$ structure where the top $(Al_{0.25}Ga_{0.75})_2O_3$ barrier is composed of a 30nm surface barrier, a 0.5 nm modulation-doped layer, and a 3nm spacer layer. (a, b) Calculation of energy band, electron concentration, and electron wave function for a structure with a modulation-doping sheet concentration of $5 \times 10^{12}$ cm$^{-3}$ and $2 \times 10^{13}$ cm$^{-3}$, respectively, on a thick $Ga_2O_3$ layer. (c) Comparison of primary channel concentration and the ratio of primary to undesired secondary channel as a function of total sheet concentration in the modulation-layer.

Similar tradeoffs are displayed in Fig. S3 comparing structures with a total sheet concentration of $5 \times 10^{12}$ cm$^{-3}$, $1 \times 10^{13}$ cm$^{-3}$, and $2 \times 10^{13}$ cm$^{-3}$. Increasing the Al mole fraction does increase the concentration in the primary channel but this effect is only significant up to a certain Al mole fraction. This Al mole fraction plateau point in primary channel charge increases for an increase in modulation-doping concentration. The more important role of the increase in Al mole fraction is deepening the triangular potential well. This increases the fraction of electrons that accumulate in the primary channel – formed at the $(Al_{0.25}Ga_{0.75})_2O_3$ / $Ga_2O_3$ interface – rather than remaining in the barrier layer.

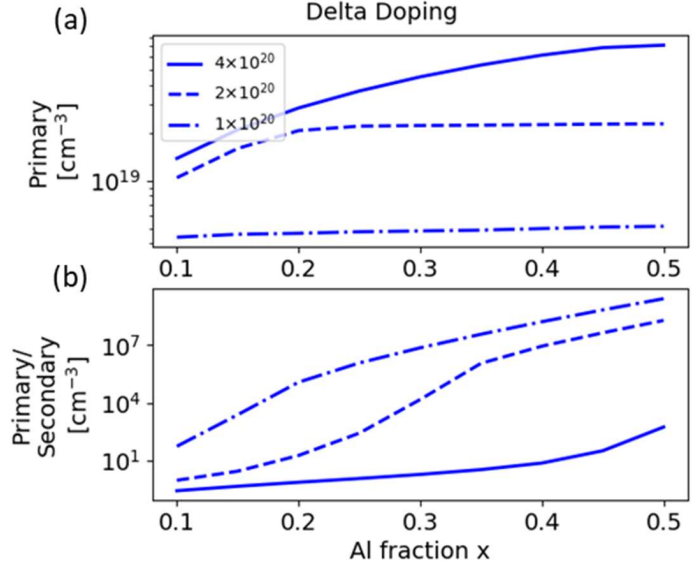

FIG. S3. (a) Carrier concentration in the primary channel and (b) ratio primary-to-secondary channel concentration for a step variation in modulation-doping concentration ($1\times10^{20}$ cm$^{-3}$, $2\times10^{20}$ cm$^{-3}$, and $4\times10^{20}$ cm$^{-3}$) as a function of Al mole fraction for a β-(Al$_x$Ga$_{1-x}$)$_2$O$_3$ / Ga$_2$O$_3$ modulation-doped structure where the top (Al$_{0.25}$Ga$_{0.75}$)$_2$O$_3$ barrier is composed of a 30nm surface barrier, a 0.5 nm modulation-doped layer, and a 3 nm spacer layer.

### C. Spacer Thickness

In the modulation-doped structure, the bottom spacer layer separates the two-dimensional electron gas in the channel from the delta doped layer. Figure S4(a) displays an (Al$_{0.25}$Ga$_{0.75}$)$_2$O$_3$ / Ga$_2$O$_3$ modulation-doped structure with a 1 nm spacer layer separating the potential well from the modulation-doped ($8\times10^{19}$ cm$^{-3}$) layer. A moderate density of electrons accumulates in the triangular potential well at the (Al$_{0.25}$Ga$_{0.75}$)$_2$O$_3$ / Ga$_2$O$_3$ interface. Closer examination of Fig. S4(a) shows that the electron wave function for the conductive channels extends to the donors in the delta-doped barrier.

In contrast, Fig. S4(b) displays a similar (Al$_{0.25}$Ga$_{0.75}$)$_2$O$_3$ / Ga$_2$O$_3$ modulation-doped structure except with a 2 nm spacer layer separating the potential well from the modulation-doped layer ($8\times10^{19}$ cm$^{-3}$). There is a slight decrease in the concentration of carriers in the primary channel; nevertheless, electron wave function in the channel does not extend to the modulation-doped donors in the (Al$_{0.25}$Ga$_{0.75}$)$_2$O$_3$ layer.

Figures S4(c) and S4(d) shows the relation of primary and secondary channel concentration over a range of spacer thickness. Across the set of modulation-doping concentration ($8\times10^{19}$ cm$^{-3}$, $1\times10^{20}$ cm$^{-3}$, and $2\times10^{20}$ cm$^{-3}$, which corresponds to total sheet concentration of $4\times10^{12}$ cm$^{-3}$, $5\times10^{12}$ cm$^{-3}$, and $1\times10^{13}$ cm$^{-3}$), there is a slight decrease in primary channel concentration for an increase in spacer thickness. At a modulation-doping concentration of $8\times10^{19}$, the primary peak channel concentration scales as $n = 2.613 \times 10^{18} e^{-0.0381 \; spacer}$. Initially increasing the spacer thickness increases the primary-to-secondary channel concentration ratio. This trend plateaus and eventually reverses for larger spacer thickness. The counteracting mechanism can be attributed to electrons in the modulation-doping layer are more isolated for a thicker spacer layer. Thus, the modulation-doped electrons more likely to accumulate at a separate potential well formed at the site of the (high) modulation-doping. In terms of peak primary-to-secondary channel concentration ratio, the optimal thickness decreases for increasing doping, e.g., at 4.5 nm for $8\times10^{19}$ cm$^{-3}$ doping (Fig. S4(c) dot-dashed line) vs 2 nm for $2\times10^{20}$ cm$^{-3}$ modulation-doping (Fig. S4(c) solid line).

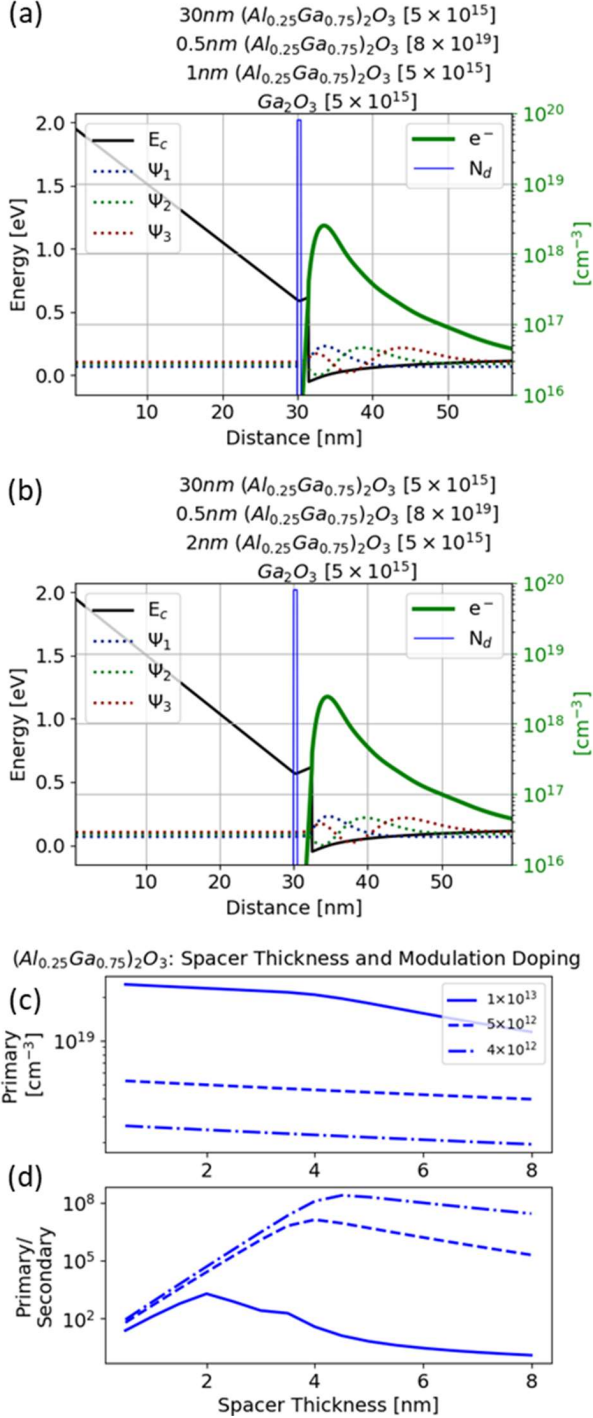

FIG. S4. Dependence on spacer thickness for a β-$(Al_{0.25}Ga_{0.75})_2O_3$ / $Ga_2O_3$ modulation-doped structure where the top $(Al_{0.25}Ga_{0.75})_2O_3$ barrier is composed of a 30nm surface barrier, a 0.5 nm modulation-doped layer, and a spacer layer. (a, b) Calculation of energy band, electron concentration, and electron wave function for a structure with a modulation-doping of $8\times10^{19}$ cm$^{-3}$ for a spacer thickness of 1 nm and 2 nm, respectively, on a thick $Ga_2O_3$ layer. (c, d) Comparison of primary channel concentration and the ratio of primary-to-secondary channel concentration for a set of modulation-doping concentration ($8\times10^{19}$ cm$^{-3}$, $1\times10^{20}$ cm$^{-3}$, and $2\times10^{20}$ cm$^{-3}$) as a function of spacer thickness. Increasing the thickness of this separation slightly reduces the concentration of electrons in the primary channel. For each modulation doping, there is an optimal thickness for maximizing the peak primary-to-secondary channel concentration.

### D. Dopant Decay

#### 1. Fixed Modulation Layer Doping

The above simulations assume a step function change in the doping profile. In this section an exponential tail of dopant atoms is added in the top barrier layer. The target doping in the modulation layer is achieved and the additional atoms in the top barrier layer arises from a memory or similar effect. An increase in decay rate constant increases the dopant atoms in the upper barrier, which increases the total dose of dopant atoms in the structure.

Figure S5(a) displays an $(Al_{0.25}Ga_{0.75})_2O_3$ / $Ga_2O_3$ modulation-doped structure with a dopant decay of 1 nm per decade. The electron profile is similar to the profile seen for the same structure with ideal step-function change in doping (see Fig. S2(a)). Electrons are only appreciably present in the triangular potential well at the $(Al_{0.25}Ga_{0.75})_2O_3$ / $Ga_2O_3$ interface. In contrast, Fig. S5(b) displays a similar $(Al_{0.25}Ga_{0.75})_2O_3$ / $Ga_2O_3$ modulation-doped structure except with a dopant decay of 5 nm per decade. The large increase in electrons across the barrier layer forms a broad secondary channel.

The exponential decay of dopant atom concentration is described by

$$N_{tail}(x) = N_0 e^{-kx},  \quad (1)$$

where $N_0$ is the doping concentration [cm$^{-3}$] in the modulation-doping layer, k is the exponential decay rate [cm$^{-1}$] and x is the position [cm] in the layer above the modulation-doped layer. The decay constant τ [nm / decade], e.g., as in Bhattacharyya et al.,[8] is converted to the exponential decay rate by $k = 1 \times 10^7 \ln(10)/\tau$. The additional charge created by the exponential decay is the integration with a solution given by

$$\sigma_{tail} = N_0 \left( \frac{1}{k} - \frac{1}{ke^{kt}} \right), \quad (2)$$

where t is the total thickness [cm] of the layer above the modulation-doped layer. The last term accounts for insufficient thickness for complete dopant incorporation. Examination of Fig. S5(b) or Table 1 shows that this tail charge (for a dopant decay constant of 5 and 10 nm/decade) is much larger than the charge in the modulation-doping layer.

| Decay [nm / decade] | Dopant Concentration [cm$^{-3}$] | Sheet Total [cm$^{-2}$] | Sheet Modulation [cm$^{-2}$] | Sheet Tail [cm$^{-2}$] |
|---|---|---|---|---|
| 0 | 8x10$^{19}$ | 4x10$^{12}$ | 4x10$^{12}$ | 0 |
| 1 | 8x10$^{19}$ | 7.5x10$^{12}$ | 4x10$^{12}$ | 3.5x10$^{12}$ |
| 5 | 8x10$^{19}$ | 2.1x10$^{13}$ | 4x10$^{12}$ | 1.7x10$^{13}$ |
| 10 | 8x10$^{19}$ | 3.9x10$^{13}$ | 4x10$^{12}$ | 3.5x10$^{13}$ |

TABLE 1. Dopant concentration in modulation layer charge in the barrier to 1×10$^{13}$. The total sheet charge is a combination of the sheet charge in the modulation layer and the tail in the top barrier layer.

Figures S5(c) and S5(d) show the relation of primary and the ratio of primary-to-secondary channel concentration over a range of dopant decay rates. Here a set of modulation-doping (8x10$^{19}$ cm$^{-3}$, 5x10$^{19}$ cm$^{-3}$, and 2x10$^{19}$ cm$^{-3}$) with lower concentration, compared to the modulation-doping in Figs. S3 and S4, is selected to offset the increase in total doping from the broad doping tail.

Figure S5(c) shows that increasing the decay rate, increases the primary channel concentration but also decreases the ratio of primary-to-secondary channel concentration. Decreasing the peak modulation-doping concentration does increase the ratio of primary-to-secondary channel concentration as observable in moving from the solid line (at 8x10$^{19}$ cm$^{-3}$) to the dashed line (at 5x10$^{19}$ cm$^{-3}$) in Fig. S5(c). Further decreasing the peak modulation-doping concentration to 2x10$^{19}$ cm$^{-3}$ (dot-dashed line in Fig. S5(c)) results in no accumulation of carriers in the primary channel - with the Fermi level well below the conduction band.

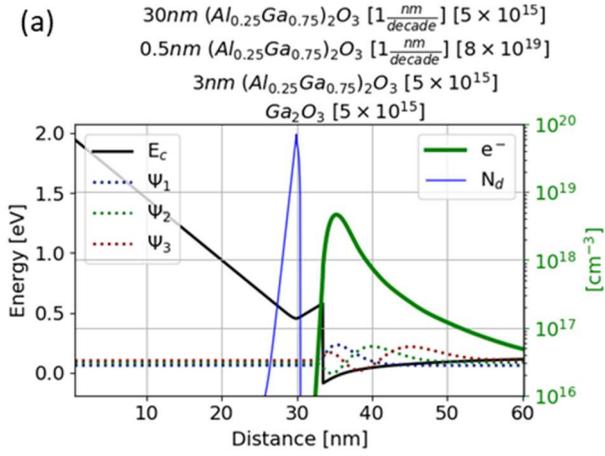
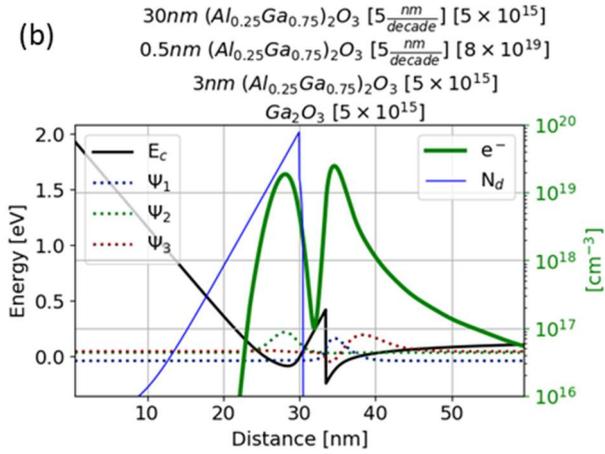
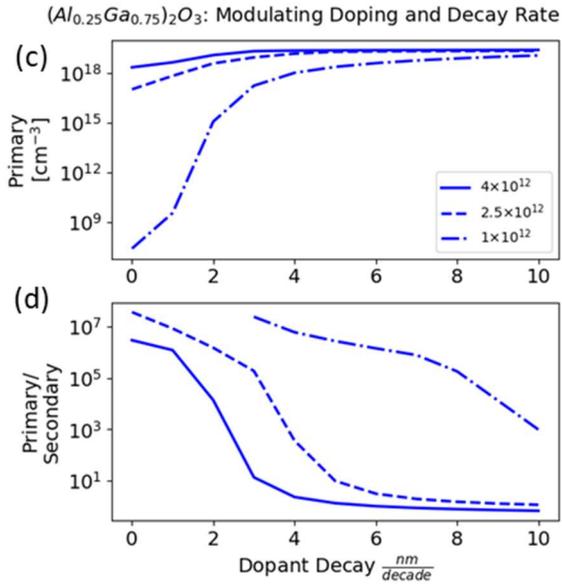

FIG. S5. Variation in dopant decay rate for a β-$(Al_{0.25}Ga_{0.75})_2O_3$ / $Ga_2O_3$ modulation-doped structure where the $(Al_{0.25}Ga_{0.75})_2O_3$ barrier is composed of a 30nm surface barrier, a 0.5 nm modulation-doped layer, and a 3 nm thickness spacer layer. (a, b) Calculation of energy band, electron concentration, and electron wave function for a structure with a modulation-doping of $8\times10^{19}$ cm$^{-3}$ for a dopant decay rate of 1 nm per decade and 5 nm per decade, respectively. (c, d) Comparison of primary channel concentration and the ratio of primary-to-secondary channel concentration for a set of modulation-doping concentration ($8\times10^{19}$ cm$^{-3}$, $5\times10^{19}$ cm$^{-3}$, and $2\times10^{19}$ cm$^{-3}$) as a function of dopant decay rate. These simulations assume that the target doping is achieved in the modulation layer and a residual doping tail step is created in the top barrier layer owing to a memory effect or similar mechanism in the deposition.

### 2. Dopant Redistribution

The more likely mechanism is that the dopant atoms in the modulation layer redistributes as an exponential tail into the top barrier layer. The target doping in the modulation layer is not achieved due to dopant preferential spreading in the direction of growth. Table 2 compares the sheet charge divided between the (reduced) sheet charge in the modulation layer and the sheet charge tail in the top barrier for a range of decay constant.

| Decay [nm / decade] | Modulation Doping [cm$^{-3}$] | Sheet Total [cm$^{-2}$] | Sheet Modulation [cm$^{-2}$] | Sheet Tail [cm$^{-2}$] |
|---|---|---|---|---|
| 0 | 1x10$^{20}$ | 5x10$^{12}$ | 5x10$^{12}$ | 0 |
| 0.5 | 7x10$^{19}$ | 5x10$^{12}$ | 3.5x10$^{12}$ | 1.5x10$^{12}$ |
| 1 | 5.4x10$^{19}$ | 5x10$^{12}$ | 2.7x10$^{12}$ | 2.3x10$^{12}$ |
| 5 | 1.9x10$^{19}$ | 5x10$^{12}$ | 9.4x10$^{11}$ | 4.1x10$^{12}$ |
| 20 | 5.44x10$^{18}$ | 4.85x10$^{12}$ | 2.72x10$^{11}$ | 4.85x10$^{12}$ |

TABLE 2. A 0.5 nm modulation-doped layer with the target doping concentration of 1x10$^{20}$ cm$^{-3}$ compared to the residual dopant concentration, which accounts for dopant redistribution. The total sheet charge is a combination of the sheet charge in the modulation layer and the exponential tail in the top barrier layer.

In the first four rows of Table 2, the total sheet charge is approximately constant because the top barrier layer is sufficiently thick to distribute the majority of the dopants in the exponential tail. At a dopant decay constant greater than 10 nm / decade a meaningful fraction of the dopant atoms will not be incorporated into the 30 nm top barrier layer.

Figure S6(a) displays an (Al$_{0.25}$Ga$_{0.75}$)$_2$O$_3$ / Ga$_2$O$_3$ modulation-doped structure with a dopant decay of 0.5 nm per decade. Electrons are only appreciably present in the triangular potential well at the (Al$_{0.25}$Ga$_{0.75}$)$_2$O$_3$ / Ga$_2$O$_3$ interface. Figure S6(b) displays a similar (Al$_{0.25}$Ga$_{0.75}$)$_2$O$_3$ / Ga$_2$O$_3$ modulation-doped structure except with a dopant decay constant of 5 nm / decade. The redistribution of charge in the top barrier layer widens the curvature of the conduction band in the barrier as expected from Poisson's equation.

Figures S6(c) and S6(d) show the relation of primary and the ratio of primary-to-secondary channel concentration as a function of dopant decay rates for a set of modulation-doping (1x10$^{13}$ cm$^{-3}$, 5x10$^{12}$ cm$^{-3}$, and 4x10$^{12}$ cm$^{-2}$) that is normalized by decreasing the doping in the modulation layer to offset the increase in dopant atoms in the broad doping tail. Compared to a structure with a step-profile in doping (decay = 0), there is a moderate decrease in electrons in the primary channel for a small increase in decay. The redistribution of the dopant atoms in the tail places the dopant atoms spatially further from the (Al$_{0.25}$Ga$_{0.75}$)$_2$O$_3$ / Ga$_2$O$_3$ interface. This effect is similar to Fig. S4 where an increase in the barrier layer thickness moderately decreases the concentration of electrons that accumulate in the primary channel potential well.

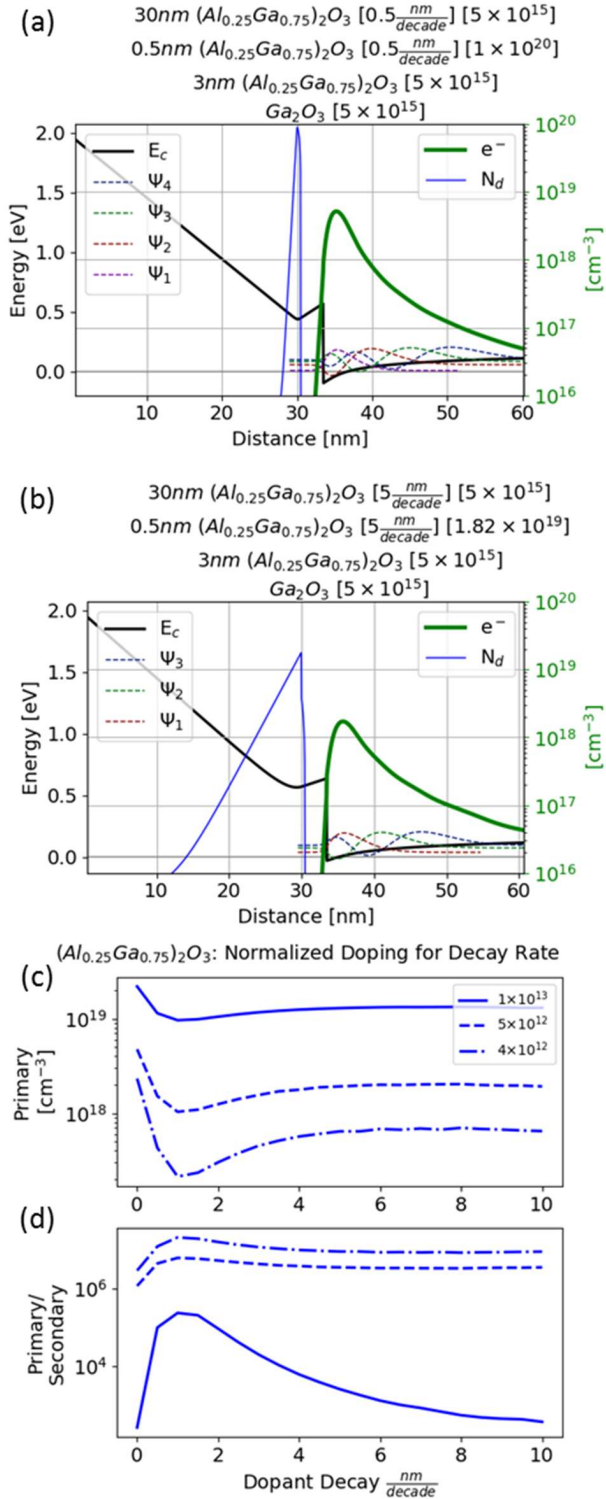

FIG. S6. Dopant redistribution in a β-$(Al_{0.25}Ga_{0.75})_2O_3$ / $Ga_2O_3$ modulation-doped structure where the $(Al_{0.25}Ga_{0.75})_2O_3$ barrier is composed of a 30nm surface barrier, a 0.5 nm modulation-doped layer, and a 3 nm thickness spacer layer. (a, b) Calculation of energy band, electron concentration, and electron wave function for a structure with a dopant decay rate of 0.5 nm per decade and 5 nm per decade, respectively. The total sheet concentration in the structure is $5\times10^{12}$ cm$^{-2}$. (c) Comparison of primary channel concentration and (d) the ratio of primary-to-secondary channel concentration as a function of dopant decay rate.

## IV. ANALYSIS

### A. Design Considerations of β-$Ga_2O_3$ MODFET

Increasing the aluminum composition in the barrier increases the conduction band offset between $(Al_xGa_{1-x})_2O_3$ and $Ga_2O_3$, which increases the accumulation of carriers in the potential well. Depositing high-quality $(Al_xGa_{1-x})_2O_3$ with an Al mole fraction greater than 0.25 has proven difficult. Still, simulations at this mole fraction (x = 0.25) show that proper selection of the barrier thickness (Fig. S4) and modulation-doping level (Fig. S2), including accounting for any asymmetric broadening (Fig. S6), can yield a dense channel confined primarily to the potential well at the $(Al_xGa_{1-x})_2O_3$ / $Ga_2O_3$ interface.

Depositing high-quality $(Al_xGa_{1-x})_2O_3$ with an Al mole fraction of 0.5 will increase in modulation-doping level that can be used without the formation of a deleterious secondary channel. The exploration of AlGaO$_3$ growth conditions, the activation energy of donors in this ultra-wide bandgap material, and the role of strain are an important area of future research.

As discussed earlier, in MOCVD of β-$Ga_2O_3$ a growth temperature dependent Si dopant decay from 50 nm/decade to 5 nm/decade was observed.[8] An asymmetric doping tail in epitaxy of semiconductors can be created by a memory effect in the growth chamber, surface-segregation, or residual atoms on the growth surface. In contrast, solid-state diffusion at the deposition temperature should produce a symmetric profile.

For the simulations displayed in Fig. S5, the total dose of dopant atoms is not normalized to account for the additional atoms in the tail. This situation occurs when additional dopant atoms supplied from the deposition environment. For example, this would occur if a deposition system had a persistent residual supply from the internal walls of the system. Deposition of β-$Ga_2O_3$ by MBE is often described as dopant or impurity atoms riding on the growth surface. This riding mechanism will create a decaying tail as the residual dopant atoms are incorporated in the subsequently deposited film.[15]

## B. Simulation Parameters for β-$Ga_2O_3$ MODFET

In this current work, the bandgap of β-$Ga_2O_3$ was set to 4.8 eV, which is the generally accepted value of the electronic bandgap.[11] Wang et al. calculated a bandgap for β-$Ga_2O_3$ and β-$Al_2O_3$ of 4.69 and 7.03 eV, respectively, with a conduction and valence offset of 2.67 eV and -0.33 eV.[12] Wang et al. predicted that corundum phase is favored for mole fraction x approaching 1 for $(Al_xGa_{1-x})_2O_3$ alloy. The conduction band offset for the $Ga_2O_3$ to $(Al_xGa_{1-x})_2O_3$ alloy in this current study uses the bowing parameters described in Wang et al.[12]

For an ALD deposited $Al_2O_3$ dielectric on β-$Ga_2O_3$, Kamimura et al. measured bandgap of 6.8 eV with a conduction and valence band offset of 1.5eV and 0.7 eV, respectively.[13] Carey et al. compared ALD deposited $Al_2O_3$ dielectric and sputter deposited $Al_2O_3$ dielectric on β-$Ga_2O_3$.[14] The ALD $Al_2O_3$ had a conduction and valence band offset of 2.23 eV and 0.07 eV; while the sputtered $Al_2O_3$ had a conduction and valence band offset of 3.16 eV and -0.86eV. The bandgap of $Ga_2O_3$ was 4.6eV and, for both the ALD and sputter deposition technique, the bandgap for $Al_2O_3$ was 6.9 eV.[14] In this current work, the bandgap for amorphous $Al_2O_3$ is 6.9 eV and the conduction band offset to β-$Ga_2O_3$ is 2.2 eV.

The surface states and Schottky barrier height influences the conduction band electron density profile. This current work assumes a negligible density of surface states. The default metal-semiconductor (MS) Schottky barrier height was set $q\phi_B$ = 1.4 eV and the default metal-semiconductor (MIS) Schottky barrier height was $q\phi_B$ = 1.5 eV. This is an approximate average of literature values for Schottky barrier heights for Ni on (010) β-$Ga_2O_3$.[15]

For reference, Bhattacharyya et al. extracted a Schottky barrier height from J-V measurements for Ni on (010) β-$Ga_2O_3$ of $q\phi_B$ = 1.27 eV and MIS $q\phi_B$ = 1.38eV ($\Delta q\phi_B$ = +0.11eV), Ni on (-201) β-$Ga_2O_3$ $q\phi_B$ = 1.04eV and MIS $q\phi_B$ = 1.21eV ($\Delta q\phi_B$ = +0.17eV), and Ni on (100) β-$Ga_2O_3$ $q\phi_B$ = 0.72eV and MIS $q\phi$ = 1.24eV ($\Delta q\phi_B$ = +0.52eV).[26] Similarly, Bhattacharyya et al., found for Schottky barrier height C-V measurements for Ni on (010) β-$Ga_2O_3$ $q\phi_B$ = 1.5eV and MIS $q\phi_B$ = 1.54eV ($\Delta q\phi_B$ = +0.04eV), Ni on (-201) β-$Ga_2O_3$ $q\phi_B$ = 1.26eV and MIS $q\phi$ = 1.28eV ($\Delta q\phi_B$ = +0.02eV), and Ni on (100) β-$Ga_2O_3$ $q\phi_B$ = 0.71eV, MIS $q\phi_B$ = 1.32eV ($\Delta q\phi_B$ = +0.61eV).[16]

Clearly, the Schottky barrier height varies depending on the metal and orientation of the β-$Ga_2O_3$ crystal. Alternative materials such as platinum group metals displayed barrier heights up to 2.1 eV.[17] Similarly, the surface preparation will alter the Schottky barrier height and the density of surface states.[18] Additionally, other works have shown the importance of the thermal design and the influence of the temperature profile on the electrical performance of the β-$Ga_2O_3$ MODFET.[19] All these issues are clearly important; nevertheless, the focus of this current supplemental is to show the dependence of the electron profile on the conduction band and doping profile in a β-$Ga_2O_3$ MODFET.

ACKNOWLEDGMENTS

The work at NRL was partially supported by DTRA Grant No. HDTRA1-17-1-0011 (Jacob Calkins, monitor) and the Office of Naval Research. The work at UF is partially supported by HDTRA1-17-1-0011. The project or effort depicted is partially sponsored by the Department of the Defense, Defense Threat Reduction Agency. The content of the information does not necessarily reflect the position or the policy of the federal government, and no official endorsement should be inferred. The work at Korea University was supported by the Korea Institute of Energy Technology Evaluation and Planning (20172010104830) and the National Research Foundation of Korea (2020M3H4A3081799).

DATA AVAILABILITY

The data that support the findings of this study are available from the corresponding author upon reasonable request.